# Dilated Balanced Cross Entropy Loss for Medical Image Segmentation

Seyed Mohsen Hosseini[1], Mahdieh Soleymani Baghshah[2]

**Abstract.** A novel method for tackling the problem of imbalanced data in medical image segmentation is proposed in this work. In balanced cross entropy (CE) loss, which is a type of weighted CE loss, the weight assigned to each class is the inverse of the class frequency. These balancing weights are expected to equalize the effect of each class on the overall loss and prevent the model from being biased towards the majority class. But, as it has been shown in previous studies, this method degrades the performance by a large margin. Therefore, balanced CE is not a popular loss in medical segmentation tasks, and usually a region-based loss, like the Dice loss, is used to address the class imbalance problem. In the proposed method, the weighting of cross entropy loss for each class is based on a dilated area of each class mask, and balancing weights are assigned to each class together with its surrounding pixels. The goal of this study is to show that the performance of balanced CE loss can be greatly improved my modifying its weighting strategy. Experiments on different datasets show that the proposed dilated balanced CE (DBCE) loss outperforms the balanced CE loss by a large margin and produces superior results compared to CE loss, and its performance is similar to the performance of the combination of Dice and CE loss. This means that a weighted cross entropy loss with the right weighing strategy can be as effective as a region-based loss in handling the problem of class imbalance in medical segmentation tasks.

## 1    Introduction

When there is a significant difference between the numbers of examples of each class in a dataset it is considered an imbalanced classification problem. The effect of imbalanced data on the performance of deep neural networks in different tasks is a well-studied area [16, 17, 28]. In segmentation tasks, the imbalance between classes can get very large, as the size of the objects can be very small compared to background. Different methods have been developed to improve the performance of the network for underrepresented classes. These methods can be divided into three main categories: data re-sampling, using region-based losses, and weighting the loss values.

---

[1] University of Tehran, smhosseini741@gmail.com
[2] Sharif University of Technology, soleymani@sharif.edu

Using region-based losses, like the Dice loss, is a popular method to improve the performance of a model trained with imbalanced data in medical segmentation tasks. A review of different loss functions for medical segmentation can be found here [1]. In region-based losses, instead of producing independent losses for each pixel of the image, a loss is calculated based on the mismatch of the segmented area and ground truth. Using a combination of different losses for imbalanced data is a common practice [1, 18]. In the combination of CE loss and Dice loss [15, 19, 24], while Dice loss provides more sensitivity for smaller objects and improves the performance for imbalanced data, CE loss leads to smoother gradient [19]. The combination of Dice and CE loss performs better than Dice loss alone [1].

Weighted cross entropy is another group of methods for solving the imbalanced data problem. The weights are usually determined based on classification difficulty [2, 3, 20], position of the pixel in the image [4], class frequency [4, 16, 22, 23], or a combination of these factors [4]. In the original U-Net [4], higher weights are given to classes based on their frequency, and also in order to have a more accurate segmentation around the boundaries of the cells, those areas are given higher weights to increase their impact on the loss. Weighted cross entropy when the weights assigned to each class are based on the inverse of class frequency is also called balanced cross entropy (BCE) [4, 16]. This method increases the impact of classes with fewer examples during training by giving them a higher weight. The performance of balanced CE in medical segmentation tasks is generally lower than CE [1, 5], and therefore it is not commonly used in popular medical segmentation tasks. The poor performance of balanced CE can also be seen in our results. Balanced CE produces lower false negatives but higher false positives and it is not a very effective method to address the class imbalance problem. It is stated in [5], that the performance of balanced CE might be improved with a better choice of weighting. In this study, a better weighting strategy is proposed and the results show that the proposed strategy can greatly improve the performance of balanced CE. Apart from the amount of dilation, the proposed weighting strategy is the same for different segmentation tasks and datasets. The performance of the proposed method on different datasets is better or similar to the performance of combination of Dice and CE loss. This shows that a weighted CE loss can be as effective as a regional loss in handling the class imbalance problem.

## 2  Method

The proposed method is based on applying balancing weights to each class and also to its surrounding area. Pixels receive a balancing weight not only based on their class but also based on their position in the image. The weights are applied to more than one class; this is different from balanced CE loss where each class receives its own specific balancing weight. Balanced CE loss is defined as:

$$L_{\text{BCE}} = \frac{-1}{N} \sum_{c=1}^{C} w_c \sum_{n=1}^{N} y_n^c \log(p_n^c) \qquad (1)$$

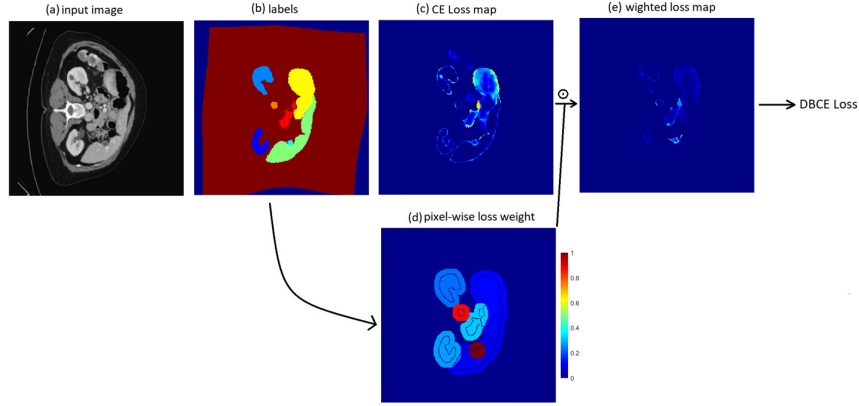

**Fig. 1.** Calculating the dilated balanced cross entropy (DBCE) loss. (a) Input image, (b) ground truth labels, (c) the cross entropy (CE) loss map before applying any weights (values are normalized for better visualization), (d) pixel-wise loss weight, which are the inverse of the area of dilated shapes of labels, the boundaries of the original labels are also included as a reference, (e) the weighted loss map, obtained by hadamard product of the loss map and loss weight (values are normalized for better visualization).

where $y^c$ and $p^c$ are the ground truth and network predictions of class c, respectively, and $w_c$ is the weight assigned to class c, defined as the inverse of class frequency. Dilated balanced CE is defined as:

$$L_{DBCE} = -\sum_{j=1}^{w}\sum_{i=1}^{h} M_{i,j} \sum_{c=1}^{C} Y_{i,j}^c \log(P_{i,j}^c) \qquad (2)$$

$$D^c = Y^c \oplus S \qquad (3)$$

$$W^c = \frac{1}{1 + \sum_{j=1}^{W}\sum_{i=1}^{H} D_{i,j}^c} D^c \qquad (4)$$

$$M_{i,j} = \max(W_{i,j}^1, \dots, W_{i,j}^C) \qquad (5)$$

where $Y^c, P^c$ are the binary ground truth mask and predicted map of each class respectively. Using the one hot labels and predicted maps, the cross entropy loss map is calculated as: $\sum_{c=1}^{C} Y_{i,j}^c \log(P_{i,j}^c)$. This loss map is depicted in Fig. 1(c). For obtaining pixel-wise loss weight map, M (Fig. 1(d)), we first use a dilation operation. In Eq. (3), the binary mask of each class is dilated using the structuring element S to produce the dilated mask of each class, $D^c$. The structuring element used in our experiments is a disk with radius R which can be treated as a hyper parameter. In Eq. (4), $D^c$ is divided by its area to produce $W^c$ which is the weighted dilated mask of class c. Finally, at every position, the maximum value of $W^c$ across all classes is chosen to produce the final pixel-wise loss weight map. The hadamard product of M and the CE loss map will give us the weighted loss map (Fig. 1(e)) and by summing that we get the final dilated balanced CE (DBCE) loss.

There are areas of the map that would be included in more than one dilated mask because of the overlap of these masks. By applying the maximum operation, those areas receive the highest weight and smaller objects will have a higher impact on the loss. In Fig. 1, by comparing the initial loss map (Fig. 1(c)) and the weighted loss map (Fig. 1(e)), it can be seen that the loss of the smallest organ and its surroundings have received the highest weight compared to larger organs.

One motivation behind assigning balancing weight to an object and its surrounding pixels is that most errors usually happen in the object and its surrounding areas. For example, in polyp segmentation the tissues around the polyp cause the most trouble for the network. The polyps usually don't have a well-defined boundary or shape and the areas around the polyps have a very similar color and texture to the polyp itself. This will lead to most errors occurring around the polyp especially in larger polyps, as it is stated in [7].

In imbalanced data, the size of the object is usually much smaller than the background, as a result in balanced CE the object gets a high weight and the background is given a much smaller weight. Thus the cost of error in the background is much smaller for the network which leads to a high false positive. The problem of false positives is more severe in areas around the object due to the general structure of the U-Net based network or any network that uses down sampling. In segmentation networks, the resolution of feature maps is usually reduced by down-sampling layers like max pooling layers. This improves the performance of the network in general but it comes at the cost of reduced spatial information. Skip connections from layers with higher resolution is used to recover spatial information but in networks with down sampling some spatial information is always lost. Low spatial accuracy, combined with relatively lower cost of error in the areas around an object, leads to a higher probability of false positives in those areas.

In the proposed method, by including the surrounding areas in the higher weight zone false positives are reduced in the areas where they are most likely to happen. Since the weight assigned to pixels is based on their spatial position as well as their class, the background pixels around an object are given much more importance and are treated differently from background pixels located far from the object.

Another motivation behind the choice of the proposed balancing method is, preventing the weights of classes with very small areas from getting too high. If the frequency of a class is very low, in other words, its area in the label map is very small and it appears in a small number of samples, its balancing weight can get very large and this will make the overall loss noisy. Dilating the label areas will ensure that their area will be above a minimum which is determined by the structuring element used for dilation. In the balanced CE, to prevent the balancing weights from getting too high the class frequency is usually calculated over the training set [1] but the performance is still lower than simple CE. In the dilated balanced CE balancing weights can be calculated in each sample, and because of the dilation, the weights will not get too high even for very small objects. By changing the area of structuring elements, we can modify how much emphasis is put on minority classes.

## 3    Experiments

To evaluate the effectiveness of the proposed loss function, multiple experiments on popular medical segmentation tasks are conducted. In all experiments, to have a more meaningful comparison between the loss functions, the training conditions, e.g. the network, training data, augmentation, learning rate, are kept the same with the only different factor being the loss function.

### 3.1    Datasets

In polyp segmentation task, the Kvasir-SEG [8] dataset is used for training and the CVC-ClinicDB [9] dataset for testing. Kvasir-SEG dataset and CVC-ClinicDB dataset include 1000 and 612 polyp images respectively. The International Skin Imaging Collaboration (ISIC) 2018 [10] is a skin lesion segmentation data set. 2594 training images and 1000 test images were used in this task. Synapse multi-organ dataset[3] from the MICCAI 2015 Multi-Atlas Abdomen Labeling Challenge contains 30 abdominal scans with 3779 slices. Following [11, 12], 18 scans are used for training and 12 scans are used for testing. The data split is the same as [12], and following [11,12], only 8 organs are segmented in this task including aorta, gallbladder, kidney left, kidney right, liver, pancreas, spleen and stomach.

### 3.2    Implementation

The implemented network is a U-Net [5] based on ResNet34 [6]. The network has 5 max pooling layers to down-sample to 1/32 of the original resolution. Deep supervision is utilized for training. Segmentation heads were added to the network at scales: 1/4, 1/8, and 1/16. A segmentation head consists of a convolution layer, bilinear up-sampling to the original resolution, and a soft-max layer. The final loss of the network is the sum of losses calculated from different scales. Like many previous studies [13, 14], the output of the network used to produce the segmentation at inference time is the up-scaled prediction map from the 1/4 scale. Weight decay of 1e-4 and the Adam optimizer with initial learning rate of 5e-4 are used. The learning rate is reduced according to a polynomial learning rate policy, (initial learning rate $\times (1 - \text{iter.}/\text{max\_iter.})^{0.9}$). All images and labels are resized to 320 × 320 for training and testing. Augmentation methods include random flipping, rotation, zooming, elastic deformation, color jittering, and Gaussian blurring. In Synapse dataset, only the dice loss of the classes that are present in each sample are considered. For polyp, skin lesion and Synapse datasets the network is trained with different losses for 600, 200 and 300 epochs respectively. The radius of the structuring element used for dilating the object area is set empirically. For the polyp and skin lesion segmentation tasks, the radius is 20 pixels and for the multi-organ segmentation task it is set to 10 pixels. The effect of the amount of dilation on performance is investigated in the results section.

---

[3] https://www.synapse.org/#!Synapse:syn3193805/wiki/217789

**Table 1.** Results of the polyp segmentation task. The network is trained on Kvasir-SEG and tested on CVC-ClinicDB. The best and second best results are highlighted in bold and underline respectively

| Loss functions | mDice | mIoU | mPrec. | mRec. |
|---|---|---|---|---|
| CE | 86.41 | 79.34 | <u>88.90</u> | 87.59 |
| Balanced CE | 84.31 | 75.99 | 81.75 | **91.95** |
| Dice + CE | <u>87.06</u> | <u>80.00</u> | **88.97** | 88.92 |
| Dilated Balanced CE (proposed) | **87.38** | **80.29** | 88.81 | <u>89.45</u> |

**Table 2.** Results of the ISIC 2018 skin lesion segmentation task. The best and second best results are highlighted in bold and underline respectively.

| Loss functions | mDice | mIoU | mPrec. | mRec. |
|---|---|---|---|---|
| CE | 88.24 | 80.72 | 84.87 | <u>95.48</u> |
| Balanced CE | 84.05 | 74.34 | 76.39 | **97.67** |
| Dice + CE | <u>88.81</u> | <u>81.58</u> | <u>86.55</u> | 94.55 |
| Dilated Balanced CE (proposed) | **88.85** | **81.62** | **86.67** | 94.43 |

The evaluation metrics are mean Dice score (mDCS), mean IoU (mIoU), mean precision (mPrec.) and mean recall (mRec.) which are averaged over the test set. In Synapse dataset, DCS and IoU of each organ is averaged over 12 test scans and then averaged over different organs to produce the overall metrics.

### 3.3 Results

The qualitative comparison results on three different datasets are presented in this section. As it can be seen in Tables 1 to 3, the performance of the proposed method is higher than CE, which shows the method can successfully alleviate the problem of imbalanced labels, in binary as well as multiclass segmentation problems. The ineffectiveness of the simple balanced CE is also obvious from the results, as it degrades the performance of CE by a great margin. But it should be noted that the balanced CE produces higher mean recall compared to other methods. This is expected as this loss increases the cost of false negatives by giving higher weights to foreground compared to background.

The combination of CE and Dice loss is a popular loss function utilized in many studies [15, 19] producing state of the art results. The results of this loss are also included in the experiments to compare the performance of the proposed loss against a top performing popular loss function in medical segmentation tasks.

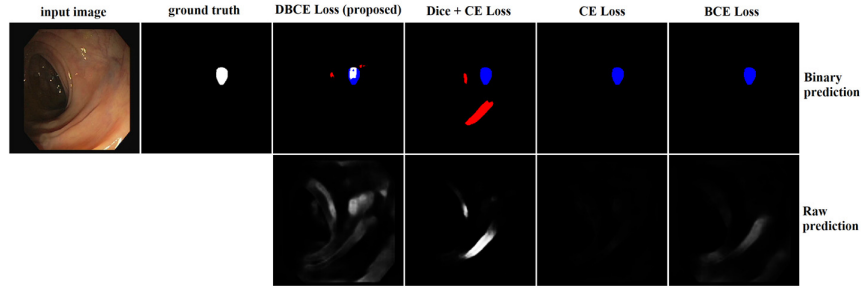

**Fig. 2.** Visual comparison of predictions obtained by 4 different losses from the polyp segmentation task. For each image the first row contains the binary predictions and the second row shows raw probability predictions. In binary predictions false positives and false negatives are marked by red and blue.

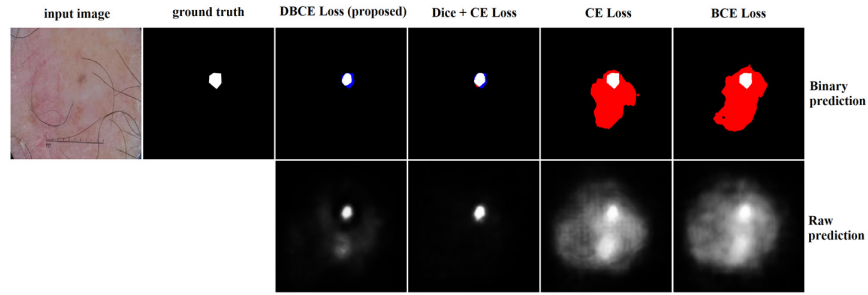

**Fig. 3.** Visual comparison of predictions obtained by 4 different losses from the ISIC 2018 skin lesion segmentation task. For each image the first row contains the binary predictions and the second row shows raw probability predictions. In binary predictions false positives and false negatives are marked by red and blue.

**Table 3.** Results of the Synapse multi-organ segmentation task. The dice score of 8 organs are also reported. The best and second best results are highlighted in bold and underline respectively.

| Loss functions | mDice | mIoU | Aorta | Gallbladder | Kidney L | Kidney R | Liver | Pancreas | Spleen | Stomach |
|---|---|---|---|---|---|---|---|---|---|---|
| CE | 81.10 | 72.30 | 90.50 | **60.65** | 83.77 | 81.23 | 95.38 | 64.94 | **91.71** | **80.66** |
| Balanced CE | 65.03 | 51.81 | 60.89 | 38.25 | 70.09 | 68.16 | 89.67 | 42.74 | 81.73 | 68.71 |
| Dice + CE | **81.75** | <u>72.74</u> | **91.79** | <u>60.48</u> | 85.54 | **83.67** | **95.49** | **68.08** | 89.79 | 79.19 |
| DBCE (proposed) | <u>81.68</u> | **72.81** | <u>90.84</u> | 59.82 | **86.63** | <u>83.28</u> | <u>95.47</u> | <u>66.54</u> | <u>91.12</u> | <u>79.70</u> |

The results of the polyp segmentation task are reported in Table 1. The network is trained with four different losses on Kvasir-SEG dataset and tested on CVC-CliniDB dataset. Balanced CE has the lowest performance except for mean recall. It can be seen

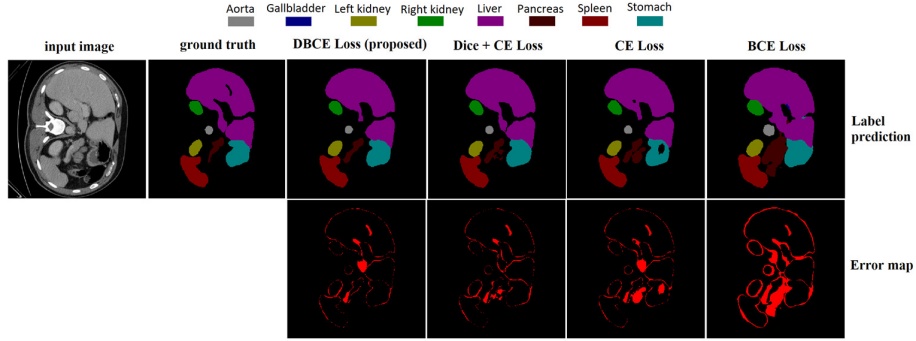

**Fig. 4.** Visual comparison of predictions obtained by 4 different losses from the Synapse multi-organ segmentation task. For each slice the first row contains the label predictions and the second row shows the error map (incorrectly labeled pixels are marked in red).

from Fig. 1 that the proposed loss can segment small areas that are missed by other losses.

The results of skin lesion segmentation task are reported in Table 2. The network is trained with four different losses on the ISIC 2018 dataset. In Table 3, the results of multi-organ segmentation task are reported. The network is trained with four different losses on 18 train scans and tested on 12 test scans. Like the previous datasets the performance of balanced CE is the lowest. In all the experiments the proposed loss outperforms the balanced CE loss and CE loss and produces similar results compared to Dice + CE loss.

It can be seen from Figures 2 to 4 that balanced CE increases the false positives compared to CE, especially in boundary pixels and areas around the objects. But errors in those areas are much lower in the results of the proposed method. The results show that the proposed weighting method can successfully address the imbalanced data problem, and improve the performance of cross entropy in different binary and multiclass medical segmentation tasks.

The amount of the dilation affects the performance of the proposed method. If the radius of the structuring element is set too low or zero the performance degrades and gets closer to the performance of the simple balanced CE. On the other hand, large dilations result in lower performances mainly because the weights would not have any balancing effect on the loss. The effect of the amount of dilation on performance for the polyp segmentation task is depicted in Figure 5. The radius of the dilation structuring element is increased from zero to 150 pixels. It can be seen that when the radius of dilation is set to zero, in other words we have a simple balanced CE, the performance is lowest. By increasing the dilation radius performance improves and then decreases for large radiuses. The best performance is achieved with radius of 20 pixels.

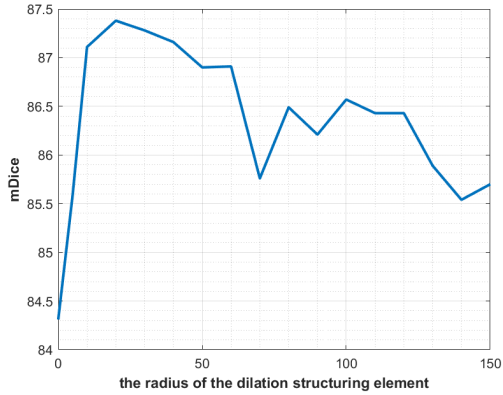

**Fig. 5.** The effect of the amount of dilation on performance for the polyp segmentation task. The radius of the dilation structuring element is increased from zero (simple balanced CE) to 150 pixels.

## 4    Conclusion

For addressing the class imbalanced problem in medical segmentation, a weighting strategy of cross entropy loss is proposed in this study. Each class and its surrounding areas receive balancing weights. Experimental results show that the proposed method can greatly improve the performance of balanced CE loss by a simple modification in its weighting method. The proposed method also outperforms the CE loss, and produces similar results compared to Dice + CE Loss. This shows that a weighted CE loss with the right weighting strategy can successfully handle class imbalanced problem in medical segmentation tasks and produce similar results compared to a region based loss.